\documentclass[12pt,a4paper]{article}

\usepackage{feynmp,amsbsy,latexsym,cite}
\usepackage{amsfonts,graphics}
\newcommand{\bc}{\begin{center}}
\newcommand{\ec}{\end{center}}
\newcommand{\no}{\noindent}

\newcommand{\Ha}{{\cal H}_{\mathrm{int}}}

\newcommand{\psib}{\bar\psi}
\newcommand{\bd}{{b^{\dag}}}
\newcommand{\dd}{{d^{\dag}}}

\newcommand{\ecd}{{\cdot}}

\newcommand{\xb}{{\boldsymbol x}}
\newcommand{\yb}{{\boldsymbol y}}

\newcommand{\pb}{{\boldsymbol p}}
\newcommand{\qb}{{\boldsymbol q}}
\newcommand{\kb}{{\boldsymbol k}}
\newcommand{\vb}{{\boldsymbol v}}
\newcommand{\rb}{{\boldsymbol r}}
\newcommand{\wb}{{\boldsymbol w}}
\newcommand{\ub}{{\boldsymbol u}}
\newcommand{\nb}{{\boldsymbol n}}
\newcommand{\lb}{{\boldsymbol l}}
\newcommand{\intx}{\int\!d^3x\;}
\newcommand{\inty}{\int\!d^3y\;}
\newcommand{\intuv}{\int\!d^3u \, d^3v\;}
\newcommand{\intrw}{\int\!d^3r \,d^3w\;}
\newcommand{\intp}{\int\!\frac{d^3p}{(2\pi)^3}\;}
\newcommand{\intq}{\int\!\frac{d^3q}{(2\pi)^3}\;}
\newcommand{\intk}{\int\!\frac{d^3k}{(2\pi)^3}\;}
\newcommand{\intpqk}{\int\!\frac{d^3p\, d^3q\,
d^3k}{(2\pi)^9}\;}
\newcommand{\R}{\mathbb{R}}

\begin{document}
\rightline{PLY-MS-99-9}
\vskip19truemm
\begin{center}{\Large{\textbf{Asymptotic Dynamics in Quantum Field Theory}}}\\ [8truemm]
\textsc{Robin Horan\footnote{email: rhoran@plymouth.ac.uk}, Martin
Lavelle\footnote{email: mlavelle@plymouth.ac.uk} and David
McMullan}\footnote{email: dmcmullan@plymouth.ac.uk}\\ [5truemm]
\textit{School of Mathematics and Statistics\\ The University of
Plymouth\\ Plymouth, PL4 8AA\\ UK} \end{center}

\begin{abstract}
\no A crucial element of scattering theory and the LSZ reduction
formula is the assumption that the coupling vanishes at large
times. This is known not to hold for the theories of the Standard
Model and in general such asymptotic dynamics is not well
understood. We give a description of asymptotic dynamics in field
theories which incorporates the important features of weak
convergence and physical boundary conditions. Applications to
theories with three and four point interactions are presented and
the results are shown to be completely consistent with the results
of perturbation theory.
\end{abstract}

\begin{quote}
\end{quote}

\medskip

\begin{quote}
\end{quote}




\section[Introduction]{Introduction}

 Descriptions of scattering in quantum field theory assume that
at large times the particles are widely separated and behave like
free particles. This assumption, which underlies the
LSZ~formalism, is incorrect for many theories. The most obvious
example of this is when the incoming or outgoing system includes
bound states, but it also fails if the physics is characterised by
long range interactions. Since most of the physics of the standard
model falls into at least one of these categories (confined
quarks, massless gauge bosons) it is very important to have a
precise understanding of the  dynamics of quantum field theories
at large times.

Generally, then, it is assumed that at asymptotic times the
Heisenberg fields become free ones
\begin{equation}\label{inout}
\lim_{t\to\infty}\phi(x)\to Z^{1/2}\phi_{\rm out}(x)\,,
\end{equation}
and similarly for $t\to-\infty$. We should note that this
behaviour can only be taken to hold as a weak limit, between
matrix elements, since otherwise (see Sect.~5-1-2 of
\cite{itzykson:1980}) one can show, from the K\"allen-Lehmann
representation, that the fields are free at all times.

The limit in (\ref{inout}) is sometimes discussed in the framework
of an \lq adiabatic approximation\rq, in which the coupling
constant is taken to be multiplied by a function which is one
during the scattering process and approaches zero for very large
(positive or negative) times. This is unsatisfactory since it
assumes the desired answer which ought rather to emerge from the
theory itself. It can also be wrong, as in the case of Quantum
Electrodynamics (QED).

QED, the paradigm for the Standard Model, has long range
interactions. The masslessness of the photon means that the
potential between static charges falls off only as $1/r$. It has
been known for a long time~\cite{Dollard:1964,Chung:1965} that this means
that (\ref{inout}) does not hold and that any attempt to impose
such a relation generates infra-red (IR) divergences in the
wave-function renormalisation constant of (charged) matter fields.

This has been studied~\cite{kulish:1970} in the relativistic
theory by Kulish and  Faddeev~(KF) and their general approach to
asymptotic dynamics has been
utilised by various
authors, see, e.g.,~\cite{Dahmen:1981nw,Nelson:1981yt,Marchesini:1984bm,Webber:1984if,
havemann:1985,Kubo:1985ns,Catani:1985xt,Contopanagos:1992yb,Pimentel:1995bi}.
We shall now give a brief sketch of the procedure adopted by KF and what their
results seem to indicate for QED.

They considered the usual QED interaction Hamiltonian
\begin{equation}\label{intham}
  \Ha(t)=-e\intx A_\mu(t,\xb)J^\mu(t,\xb)\,,
\end{equation}
where $J^\mu(t,\xb)=\psib(t,\xb)\gamma^\mu\psi(t,\xb)$ is the
(conserved) matter current. In order to carry out the LSZ
reduction of the $S$-matrix we must be in the interaction picture
so that, although the time evolution of the states is determined
by (\ref{intham}), the evolution of the fields themselves is given
by the free Hamiltonian. One may then insert the free field
expansions in (\ref{intham}). These plane wave expansions are
\begin{equation}\label{psi_free}
\psi(x)=\intp\frac1{\sqrt{2E_{\smash{p}}}}\left\{
b(\pb,s)u^s(p)e^{-ip\ecd x}+ \dd(\pb,s)v^s(p)e^{ip\ecd x} \right\}
\end{equation}
where the notation implies a sum over the $s$ indices. Working in
Feynman gauge we have
\begin{equation}\label{a_free}
A_\mu(x)=\intk\frac1{2\omega_{\smash{k}}}\left\{
a_\mu(\kb)e^{-ik\ecd x}+ a_\mu^\dag(\kb)e^{ik\ecd x} \right\} \,,
\end{equation}
where $k_0=E_p=\sqrt{|\pb|^2+m^2}$ and $\omega_k=|\kb|$. Inserting
these into (\ref{intham}) results in eight terms which may be
grouped according to the positive and negative frequency
components of the fields. Each of these pieces will have a time
dependence of the form $e^{i\psi t}$ where $\psi$ involves sums
and differences of energy terms.

KF claimed that, for $|t|\to\infty$, only terms with $\psi$
tending to zero contribute to the asymptotic dynamics. Since the
spatial integration in (\ref{intham}) generates a momentum delta
function, only four terms with  $ \psi=\pm
(E_{p+k}-E_p\pm\omega_k) $ would then have a large $t$-limit. This
vanishing of $E_{p+k}-E_p\pm\omega_k\approx0$ can only take place
in QED because the photon is massless, and it only occurs for soft
photons, i.e., for $\omega_k\approx0$. This is in accord with
perturbation theory: the breakdown of the $S$-matrix occurs for
soft photons and giving the photon a small mass acts as a cut-off
on these divergences.
An asymptotic approximation to (\ref{intham}) is obtained from the
lowest order term of the Taylor expansion, in powers of $k$, of
the Hamiltonian. This yields
\begin{equation}\label{hint_as}
\Ha^{\mathrm{as}}(t)=-\mathrm{e}\intx
A_\mu(t,\xb)J^\mu_{\mathrm{as}}(t,\xb)
\end{equation}
where
\begin{equation}\label{j_as}
  J^\mu_{\mathrm{as}}(t,\xb)=\intp\frac{p^\mu}{E_p}\rho(p)\delta^3
  \biggl(\xb-\frac{\pb}{E_p} t\biggr)\,,
\end{equation}
and $\rho(p)$ is the charge density
\begin{equation}\label{chg}
  \rho(\pb)=\sum_{s}\Bigl(\bd(\pb,s)b(\pb,s)-\dd(\pb,s)d(\pb,s)\Bigr)
\,.
\end{equation}
According to KF therefore, the asymptotic Hamiltonian is the
integral over all momenta of the current associated with a charged
particle of velocity $p^\mu/E_p$. This non-vanishing Hamiltonian
finds its perturbative expression in the branch cuts, rather than
poles, in the on-shell Green's functions in the matter fields of
QED. An attractive aspect of this theory is that it completely
dispenses with the adiabatic approximation. Although this
discussion seems to pick up the problems in applying the
LSZ~scheme to QED and, in particular, correctly identifies the
problem with long wavelength  ($\omega_k\approx 0$) photons, it
cannot be regarded as the end of the story.

These arguments are used extensively in other theories, such as
QCD, where the physics is not well understood, and where greater
reliance is put on the mathematics. On the other hand the theory
is employed at the level of operators whereas it is more
appropriate, in quantum field theories,  to work at the level of
matrix elements and weak limits. The KF approach also makes no
connection between the large time limit and the separation of
particles at large distance.
 As we shall see below, the naive application of this approach to
massive $\phi^4$ theory would indicate that the LSZ~formalism
should also break down, which it very evidently does not.

This paper is concerned with constructing a new approach to
asymptotic dynamics within the context of weak convergence, and
with appropriate physical boundary conditions corresponding to the
separation of particles.  We will apply it to a variety of
interactions and show that it yields results which are completely
consistent with what is known from explicit perturbative
calculations.

In Sect.~2 we will show that the KF~argument cannot be applied to
four point interactions. Sect.~3, which is the heart of this
paper, develops the method for massive~$\phi^4$ theory and then
applies it to both the three and four point interactions of scalar
QED. In this way we will see both the well known spin independence
of the IR~structure in the abelian theory with massive charges
and, for the three point vertex, regain the results
of~\cite{kulish:1970}. A discussion of the implications of these
results for perturbative calculations in the standard model is
presented in Sect.~4. Some technical details are given in
appendices.

\section[Four Point Interactions]{Four Point Interactions}

 As we have indicated in the introduction, there are objections
to the KF view of asymptotic dynamics: firstly, that the
statements have been framed specifically for the picture of the
$\omega_k\approx 0$ infra-red theory, which is well known to
suffer from divergences, and not as a general statement on the
behaviour of asymptotic limits; secondly, and more critically,
that the prescription does not translate to other quantum field
theories. It is the latter objection which is the most serious and
the one which we shall now demonstrate.

 To be precise, we shall show that the KF
argument applied to the case of massive  $\phi^4$ theory is not
sufficient to show that the asymptotic limit of the coupling term
vanishes. However, $\phi^4$ theory is the standard textbook
example of an interacting quantum field theory and its
perturbation theory is straightforward -- the coupling must vanish
for well separated particles .

 To begin then, let us take the standard free field
expansion of the scalar field
\begin{equation}
\label{fe1}
 \phi(\xb )= \intk \frac{1}{2E_k}\left(a(\kb)e^{-ik\cdot
x}+a^\dag(\kb)e^{ik\cdot x}\right) \,,
\end{equation}
with $E_k=\sqrt{|\kb|^2+m^2}$\, and $m$ the mass of the particle.
The commutator relations are
\begin{equation}
\label{com1}
[a(\kb)\,,a^\dag(\kb')]=(2\pi)^3\,2E_k\,\delta^3(\kb-\kb')
\end{equation}
and the interaction part of the Hamiltonian for the theory under
consideration is then
\begin{equation}
\label{ham1} \Ha=\frac{\lambda}{4!}\intx :\,\phi ^4
(\xb\,,t)\,:\,,
\end{equation}
where the : : indicates normal ordering.

When the expansion (\ref{fe1}) is inserted into (\ref{ham1}) and
the resulting expression is simplified, then, after normal
ordering, it will be found to consist of twelve terms, each of
which has an exponential term where the exponent is made up of
sums and differences of the energy eigenvalues. Some of these
exponents are obviously non-vanishing. According to the methods of
KF which were described in the introduction, the integrals
containing these exponentials may be ignored. However, not all of
the integrals involved have exponents which are so easily dealt
with and one of these, which we shall now consider, is
\begin{equation}
\label{trm1}
 \intpqk
\frac{a^{\dag}(\kb)a^{\dag}(\pb+\qb-\kb)a(\pb)a(\qb)}{2E_p2E_q2E_k2E_{p+q-k}}
e^{-it(E_p+E_q-E_k-E_{p+q-k})}\,.
\end{equation}
Applying the methods described in the introduction, any
non-vanishing asymptotic dynamics will come from those terms and
those momenta for which the exponent vanishes. To this end we must
determine if the equation
\begin{equation}
\label{exp1}
 E_p+E_q-E_k-E_{p+q-k}=0
\end{equation}
has any solutions.

Far from this being a difficult problem, upon reflection it
becomes obvious that there are infinitely many solutions! This is
most easily seen by noting that the problem is equivalent to that
of finding solutions to the simultaneous system of equations
\begin{eqnarray}
\label{exl}
\sqrt{|\pb |^2+m^2}+\sqrt{|\qb |^2+m^2}&=&\sqrt{|\kb
|^2+m^2}+\sqrt{|\lb |^2+m^2}\nonumber\\
\pb + \qb &=&\kb +\lb \,.
\end{eqnarray}
The incoming and outgoing momenta must have some connection with a
scattering process so that the obvious, trivial solutions, which
can be found by taking, say, $\qb =\lb =\boldsymbol{0}$, will be
ignored.
\begin{figure}[tbp]
\begin{center}
\includegraphics{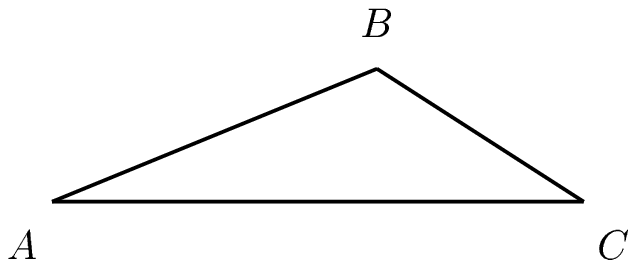}
\end{center}
\caption{Triangle of Momenta}
        \label{vecs1}
\end{figure}

 Referring to Fig.~\ref{vecs1}, let $\pb=\stackrel{\longrightarrow}{AB},$ and
$\qb=\stackrel{\longrightarrow}{BC}\,.$ Now imagine pivoting the
rigid triangle $ABC$, with the line $AC$ as the hinge, to get a
new triangle which is congruent to $ABC\,.$ If the new triangle is
$AB'C$ then take $\kb=\stackrel{\longrightarrow}{AB'}$ and
$\lb=\stackrel{\longrightarrow}{B'C}\,.$ The vectors
$\pb,\qb,\kb,\lb$ then automatically satisfy both the conditions
of (\ref{exl}).

 One can do this with any two non-parallel vectors, for example
(with $m=1$) take
$\pb=(1,1,0),\qb=(1,-1,0),\kb=(1,1/\sqrt{2},1/\sqrt{2}),
\lb=(1,-1/\sqrt{2},-1/\sqrt{2})\,.$ Another example is
$\pb=(-4/\sqrt{3}, 2\sqrt{2}/\sqrt{3},0), \qb={\boldsymbol
0},\kb=(-2/\sqrt{3},\sqrt{2}/\sqrt{3},1),$ and
$\lb=\pb+\qb-\kb\,.$

 No physical meaning could be given to such an
arbitrary set of momenta. Following the arguments of KF however,
we conclude that the existence of this wide range of momenta for
which (\ref{exp1}) vanishes, indicates that the asymptotic
dynamics of the system is determined by this set of points. This
would suggest that problems could arise in the associated
perturbation theory for massive $\phi^4$ theory, and this is well
known to be false for this textbook example of a quantum field
theory.

 At this point we wish to stress that this apparent
contradiction has implications beyond this simple example as there
are many important quantum field theories, such as QCD and the
Higgs mechanism, in which four point interactions have a major
role. If there is a deficiency in the understanding of the
asymptotic dynamics in this simple scalar four point theory then
it is difficult to see how one might proceed on this basis,  with
any confidence, in other theories of the standard model. It is
clear from this that the KF argument needs considerable refinement
if it is to be applied to the standard model.

\section[General Approach to Asymptotic Dynamics]{General Approach to Asymptotic Dynamics}

The major flaw in the KF argument is its reliance on strong
operator convergence. Instead of concentrating on the Hamiltonian
itself, we shall focus our attention on matrix elements.  The
limits we shall consider are weak limits and this is in keeping
with the LSZ formalism in quantum field theories.

 If $\Psi_{\mathrm{IN}}$ represents an incoming wave packet in this
scattering experiment and $\Psi_{\mathrm{OUT}}$ is an outgoing
wave packet, then the matrix element of interest is
$<\!\!\Psi^\dagger_{\mathrm{OUT}}|\Ha|\Psi_{\mathrm{IN}}\!\!>$.
This is a time dependent c-number and the elementary notion of
convergence may be adopted in the investigation of its asymptotic
limit. In the case of massive scalar $\phi ^4$ it will be shown
that, under conditions which have a straightforward physical
interpretation, its asymptotic limit is zero.

 This section is organised as follows. In Sect.~\ref{phi} we
shall use massive $\phi ^4$ theory as a vehicle to study
asymptotic dynamics. Having established the basic method and shown
that it is consistent with what is already known for $\phi^4$
theory, we shall then employ it to determine the asymptotic
dynamics of scalar QED.

 There is a particularly interesting feature of scalar QED
which makes the study of its asymptotic dynamics worthwhile. The
interaction Hamiltonian in scalar QED consists of a sum of two
terms, a three point interaction term, similar to that in
(fermionic) QED, and a four point interaction term, which has no
parallel in QED, so that the dynamics of scalar QED is richer than
that of fermionic QED. From perturbation theory it is known that
(fermionic) QED and scalar QED have the same infra-red problems,
so their {\it asymptotic} dynamics are the same. Perturbation
theory also tells us that the four point interaction term is
divergence free so it must have vanishing asymptotic dynamics. We
would expect this to emerge, in a natural manner, from any
satisfactory theory which describes the asymptotic dynamics of
scalar QED.  Further, since the infra-red problem is spin
independent, the asymptotic dynamics of the three point
interaction will, with the obvious changes, be the same in either
fermionic or scalar QED.

We shall begin our examination of scalar QED in
Sect.~\ref{quartic} with the quartic term, which turns out to be
the easier to deal with. We shall show that its asymptotic
dynamics is similar to that of the scalar $\phi^4$ theory and has
a zero limit. After this, in Sect.~\ref{cubic}, we shall turn our
attention to the cubic term and the infra-red problem. We shall
prove that the asymptotic dynamics for the infra-red problem is
exactly the same as that for a system in which the Hamiltonian is
derived from a current associated with a moving charged particle
with known, non-trivial asymptotic dynamics.

\subsection{Scalar $\phi^4$ Theory}
\label{phi}

 Consider the following incoming and outgoing wave packets.
\begin{eqnarray}
\label{pks1}
\Psi _{\mathrm{IN}}&=&\intrw f(\rb)g(\wb)a^\dag (\rb)a^\dag
(\wb)|0\!\!>\,, \nonumber\\
&&  \\
\Psi _{\mathrm{OUT}}&=&\intuv h(\ub)i(\vb)a^\dag (\ub)a^\dag
(\vb)|0\!\!>\,,\nonumber
\end{eqnarray}
with the functions $f,g,h,i$ being test functions for the wave
packets. Referring to (\ref{fe1}), (\ref{com1}), (\ref{ham1}), and
the expressions (\ref{pks1}), one finds that
$<\!\!\Psi^\dagger_{\mathrm{OUT}}|\Ha|\Psi_{\mathrm{IN}}\!\!> $
reduces to a single integral which is proportional to
\begin{equation}
\label{trm2} \int\!d^3p\, d^3q\, d^3k\, h(\kb )i(\pb +\qb
-\kb)f(\pb )g(\qb )e^{-it\psi}\,,
\end{equation}
with the exponent $\psi$ having the value
$\psi=E_p+E_q-E_k-E_{p+q-k}\,.$

 Notice that the exponent in this term has essentially the same structure as
the term in (\ref{trm1}) which appeared to cause problems when
applying the methods of KF. The difference now is that
(\ref{trm2}) is a straightforward integral and not an operator, so
that elementary methods may be applied to find the asymptotic
limit. The machinery we shall employ is the {\it method of
stationary phase} \cite{Guillemin:1977}. Briefly, this says that,
provided there is no point in the region of integration at which
all of the first order partial derivatives of $\psi$ are zero,
then the integral (\ref{trm2}) tends to zero as $|t|\rightarrow
\infty\,.$

 The terms in $\psi$  have the form $E_l=\sqrt{|\lb |^2+m^2}$
and, since $m\neq0,$ they will all have first-order partial
derivatives for all values of $\lb.$ The first order derivatives
of $\psi $ are then given by
\begin{eqnarray}
\label{drv1}
\frac{\partial \psi}{\partial p_i} &=&
\frac{p_i}{E_p}-\frac{p_i+q_i-k_i}{E_{p+q-k}}\nonumber \\
&&\nonumber\\
\frac{\partial \psi}{\partial q_j}
&=&\frac{q_j}{E_q}-\frac{p_j+q_j-k_j}{E_{p+q-k}} \\
&&\nonumber\\
 \frac{\partial \psi}{\partial k_n} &=&
 \frac{k_n}{E_k}-\frac{p_n+q_n-k_n}{E_{p+q-k}}\,.\nonumber
\end{eqnarray}

If at some point all of these are zero then, in particular,
$\partial\psi/\partial p_i=\partial\psi/\partial q_j=0$ for all
possible values of $i,j\,.$ This implies that $\pb /E_p=(\pb +\qb
-\kb )/E_{p+q-k}$ and $\qb/E_q=(\pb +\qb -\kb )/E_{p+q-k}$ so we
must also have $\pb /E_p=\qb/E_q\,.$ If we can exclude the set of
points for which this condition holds from the domain of
integration, then the integral in (\ref{trm2}) will vanish as
$|t|\rightarrow\infty\,.$

 In (\ref{trm2}), the test functions $f,g$ for the the incoming wave
packet have the variables $\pb,\qb$ as arguments. The expressions
$\pb /E_p$ and $\qb/E_q$ represent the velocities of the
respective incoming wave packet. Experimentally, scattering is
prepared by setting up the apparatus in such a way that the two
beams of particles are brought together from different directions,
i.e., with different velocities. This information can be
incorporated into the incoming wave packet by ensuring that the
supports of the test functions exclude the possibility that $\pb
/E_p=\qb/E_q\,.$ The precise statement of the requirement is that
{\it  the test functions $f,g$ must have non-overlapping supports
in velocity space.}  This condition on the test functions, of
having non-overlapping supports in velocity space, is central to
the construction of the $S$-matrix (see Sect.~13.4 of
\cite{Glimm:1987ng}).

To restate, if the test functions $f,g$ have non-overlapping
supports in velocity space then the integral in (\ref{trm2})
vanishes as $|t|\rightarrow\infty\,.$  This is exactly the
behaviour that one would expect for this particular scattering
process but a further question remains to be answered: what
constraints does this choice of  test functions for the incoming
wave packet impose on the outgoing wave packet? If this picture is
to display all of the features of this particular scattering
process then some combinations of outgoing particles must be
excluded. The outgoing particles must behave as free particles at
asymptotically large times which means that, as was the case for
the incoming wave packets, their test functions must also have
non-overlapping support in velocity space. However, this must be a
consequence of the condition imposed on the test functions for the
incoming wave packet and not an independently imposed condition.

 The arguments of the test functions for the outgoing wave
packet are  $\pb+\qb-\kb$ and $\kb,$ and the equality of these two
variables is equivalent to $\pb+\qb=2\kb$. This is simply the
expression of conservation of momentum. Another principle in any
scattering theory is conservation of energy. In this case this is
expressed as $E_p+E_q=2E_k\,.$ Finally then, we must show that the
two conditions
\begin{eqnarray}
\label{con1}
\pb +\qb &=&2\kb\, , \nonumber\\
E_p+E_q&=&2E_k \,,
\end{eqnarray}
are incompatible with the functions $f,g$ having non-overlapping
support in velocity space. Since the masses of the two incoming
particles are equal, the equation $\pb /E_p=\qb/E_q$ is equivalent
to $\pb =\qb$ and non-overlapping in velocity space is equivalent
to non-overlapping in momentum space. In this case {\it we need to
show that if} $\pb\neq\qb$ {\it then the conditions of}
(\ref{con1}) {\it are impossible}.

 Again, for simplicity, let us take the mass $m=1\,.$ The first
equation in (\ref{con1}) means that the vectors $\pb,\qb,\kb,$ are
coplanar. In that case, we may choose a unit vector $\nb $
orthogonal to this plane and write $\pb '=\pb+\nb,\, \qb
'=\qb+\nb,\,\kb '=\kb+\nb\,.$ Since $\nb$ is orthogonal to the
plane of $\pb,$ we have $|\pb '|=\sqrt{|\pb |^2+1}=E_p,$ with
similar expressions for $\qb,\kb\,.$ This means that (\ref{con1})
may now be written as
\begin{eqnarray}
\label{con1a}
\pb '+\qb '&=&2\kb '\nonumber\\
|\pb '|+|\qb '|&=&2|\kb '|\,.
\end{eqnarray}
From the triangle inequality we know that this is only possible
when $\pb'$ and $\qb '$ are parallel, i.e. if there is a number
$\lambda$ such that $\pb '=\lambda\qb '\,.$ In terms of $\pb,\qb$
and $\nb,$ this can be rearranged into the form $\pb -\lambda\qb
=(1-\lambda)\nb\,.$ The only solution for this is with $\lambda=1$
so that $\pb =\qb\,.$

There are other possible matrix elements that are associated with
the four point interaction term. There is the possibility that the
incoming wave packet consists of a single field, with the outgoing
wave packet made up of three fields, and there is the reverse.
Both of these cases may be treated in precisely the same manner as
the above case, and with precisely the same conclusions. We omit
the details.

\subsection{Scalar QED: the four point interaction.}
\label{quartic}
The interaction in scalar QED is more complicated than that of QED
due to the existence of the extra term representing a four point
interaction. In this section we shall study the asymptotic
properties associated with this term and show that it has trivial
asymptotic dynamics. This will require extending our techniques
since we shall have to deal with wave packets which have massless
particles as an essential part of their structure.

 The method of stationary phase that was used to determine the
limit of (\ref{trm2}), in Sect.~\ref{phi}, was dependent upon some
of the properties of the partial derivatives (\ref{drv1}) of the
various energy eigenvalues. In the case of {\it massless}
particles the energy eigenvalues, which are of the form
$\omega_k=|\kb|,$ are not differentiable at the origin but this
will not be a barrier to the application of this technique.

We begin by writing out the full, normal ordered interaction
Hamiltonian for scalar QED, which is
\begin{equation}
\label{ham2} \Ha(t) =-e\intx :J^\mu(\xb )A_\mu (\xb ):
\end{equation}
with : : being normal ordering and where the current $J^\mu $ is
given by
\begin{eqnarray}
\label{cur1}
 J^\mu&=&i(\phi^\dag \partial^\mu\phi
-\partial^\mu\phi^\dag\phi)-e g^{i\mu}A_i\phi^\dag\phi \nonumber\\
%
%
&=& i J_1^\mu -e  J_2^\mu  \\
%
%
&=&i\ (J_{11}^\mu+J_{12}^\mu)-e J_2^\mu \,,\nonumber
\end{eqnarray}
with the obvious meaning given to the components of $J^\mu$
defined in (\ref{cur1}).

 We shall work in Feynman gauge and take the plane wave
expansions given by
\begin{eqnarray}
\label{fe2}
\phi(\xb )&=& \intp \frac{1}{2E_p}\left(a(\pb)e^{-ip\cdot
x}+b^\dag(\pb) e^{ip\cdot x}\right)\nonumber\\
%
%
&=&\phi_+(\xb )+\phi_-(\xb ) \nonumber\\
&&\\
A_\mu(\xb )&=&\intk \frac{1}{2\omega_k}(a_\mu(\kb)e^{-ik\cdot
x}+a_\mu^\dag(\kb) e^{ik\cdot x}) \nonumber\\
%
%
&=& A^+_\mu(\xb )+A_\mu^-(\xb )\,. \nonumber
\end{eqnarray}
 The commutator for the photon is
given by
\begin{equation}
\label{com2}
[a_\mu(\pb),a^\dag_\nu(\qb)]=-(2\pi)^32\omega_kg_{\mu\nu}\delta^3(\pb-\qb)\,.
\end{equation}

When (\ref{fe2}) is substituted into (\ref{ham2}) and rearranged,
the quartic interaction will be found to consist of 12 terms, each
of them with differing exponential terms corresponding to the
different possible interactions. We shall consider the matrix
element represented by the following diagram.
\begin{center}
$$
\includegraphics{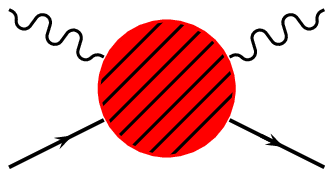}
$$
\end{center}
\no We shall take as our wave packets the following expressions
\begin{eqnarray}
\label{pck2}
\Psi_{\mathrm{IN}}&=&\intrw f(\rb )b^\dag (\rb )\,c^\nu(\wb
)a^\dag_\nu(\wb )|0\!\!>\,, \nonumber\\
&&\\
\Psi_{\mathrm{OUT}}&=&\intuv g(\ub )b^\dag(\ub )\,h^\mu (\vb
)a^\dag _\mu (\vb )|0\!\!>\,, \nonumber
\end{eqnarray}
 where the $f,c^\nu,g,h^\mu$ are the respective test functions. One
then finds that the amplitude
$<\!\!\Psi^\dag_{\mathrm{OUT}}\,|\,\Ha(t)|\,\Psi_{\mathrm{IN}}\!\!>$
is a single term, constructed from $J_2^\mu,$ which is an integral
proportional to
\begin{equation}
\label{trm3}
 \int  d^3p\, d^3q\, d^3k\, c^i(\pb )h_i(\qb )g(\kb )f(\kb +\qb -\pb
)e^{i\psi t},
\end{equation}
 where $\psi=\omega_p +E_{k +q -p}-E_k-\omega_q\,.$

 Notice that the Einstein summation convention means that this
integral is actually a sum of three integrals, with $i=1,2,3.$ The
incoming and outgoing charged fields must be separated, i.e., the
test functions $f$ and $g$ must have disjoint support. If not,
then, by conservation of momentum, there will be no separation of
the incoming and outgoing photons. In that case, no scattering
will have taken place.  Thus it must follow that, {\it for each}
$i$, the test functions $c_i,h_i$ must have disjoint support so
that at least one of them will not have the zero vector in its
support. Without loss of generality, let us suppose that this is
$c_i.$

 The function $\psi$ will then have continuous partial
derivatives in $\pb ,\kb$ and these are given by
\begin{eqnarray}
\label{ders2}
\frac{\partial \psi}{\partial
p_i}&=&\frac{p_i}{\omega_p}-\frac{k_i+q_i-p_i}{E_{k+q-p}}
\nonumber\\
&&\\
\frac{\partial\psi}{\partial
k_j}&=&\frac{k_j+q_j-p_j}{E_{k+q-p}}-\frac{k_j}{E_k}\,.\nonumber
\end{eqnarray}
 The vanishing of these expressions  for all $i,j$ would then
imply that $\pb /\omega_p=\kb /E_k$
 which is impossible, since
the former is a unit vector while the latter is not. The method of
stationary phase (applied only to the variables $\pb ,\kb $) can
again be used to prove that (\ref{trm3}) will vanish as $|t|
\rightarrow\infty\,.$

 There are other possible scattering events that are described
by this four point interaction Hamiltonian, e.g.,  when the wave
packets consist of two incoming photons and two outgoing charges
(or vice versa). If the photons are separated according to our
scheme (their test functions have disjoint support),  then a
similar exercise will show that this picture also gives rise to
vanishing asymptotic dynamics.

\subsection{Scalar QED: the infra-red approximation}
\label{cubic}
 The cubic term in the interaction Hamiltonian (\ref{ham2}),
which comes from the $J_1^\mu$ term in (\ref{cur1}), is made up of
two parts, $J_{11}^\mu$ and $J_{12}^\mu\,.$ These two terms are
similar in their structure and both contribute to the problem of
infra-red divergences. We shall consider only the first of these,
$J_{11}^\mu,$ the results for $J_{12}^\mu$ being substantially the
same.

 As in QED, the infra-red problem in scalar QED occurs in
relation to a scattering process in which an incoming charged
particle emits a photon.
\begin{center}
$$
\includegraphics{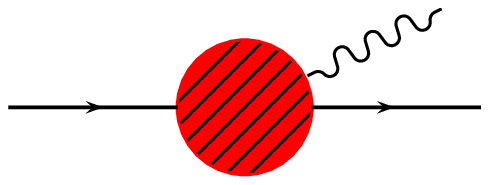}
$$
\end{center}
 We shall consider the
case when the wave packets are given by
\begin{eqnarray}
\label{pck3}
\Psi_{\mathrm{IN}}&=&\inty f(\yb )b^\dag(\yb) |\,0\!\!>
\nonumber\\
\Psi_{\mathrm{OUT}}&=&\intuv g(\ub )b^\dag(\ub)\,h^\mu(\vb
)a^\dag_\mu(\vb)|\,0\!\!>\,.
\end{eqnarray}
The matrix element will then be found to be given by
\begin{equation}
\label{trm4}
<\!\!\Psi^\dag_{\mathrm{OUT}}\,|\,\Ha(t)|\,\Psi_{\mathrm{IN}}\!\!>=-e
\int d^3 q\, d^3k
 f(\qb +\kb )g(\qb )q^\mu h_\mu(\kb ) e^{-i\psi t}
\end{equation}
where $\psi = E_{q+k}-E_q-\omega_k\,.$ Now it is easy to see why
the previous method \emph{cannot} be applied in this case. Since
the photon is massless, the corresponding energy eigenvalue is
$\omega_k=|\kb |$. The expression for $\psi$, therefore,  will not
have partial derivatives in $k_i$ at $|\kb| =0,$ and the partial
derivatives in $q_j$ will vanish when $|\kb|=0,$ for any value of
$\qb.$ It is this regime, when $|\kb| =0$, which gives rise to the
problem of infra-red divergences and the difficulties associated
with its asymptotic dynamics cannot be avoided.

Following KF  we shall compare the asymptotic dynamics of this
scattering process with that governed by the scalar version of the
asymptotic Hamiltonian (\ref{hint_as}).

The system that we shall take for our comparison is then the one
defined the {\it asymptotic} current given by
\begin{equation}
\label{cur2}
J^\mu_{\mathrm{as}}(\xb )=i\intq \left(
\frac{1}{2E_q}\right)^2b^\dag (\qb)b(\qb)q^\mu \delta^3 ( \xb
-\frac{\qb }{E_q}t)\,,
\end{equation}
where we omit terms which do not contribute to this matrix
element. The interaction Hamiltonian is given by
\begin{equation}
\label{ham3}
 \Ha ^{\mathrm{as}} =-e\intx J^\mu_{as}(\xb )\,A_\mu (\xb )\,.
\end{equation}

 If we take the incoming and outgoing wave packets
(\ref{pck3}) and contract, in the usual fashion, with
$\Ha^{\mathrm{as}}(t)$ then we obtain the amplitude
\begin{equation}
\label{term5}
 <\!\! \Psi_{\mathrm{OUT}}^\dag\,|\Ha^{
\mathrm{as}}(t)|\,\Psi_{\mathrm{IN}}\!\!>=-e \int d^3 q \,d^3k
f(\qb )g(\qb )q^\mu h_\mu(\kb ) e^{-i\psi 't}\,,
\end{equation}
 where $\psi '=\qb\cdot\kb/E_q-\omega_k\,.$

 The asymptotic dynamics of the systems defined by the
interaction Hamiltonians (\ref{ham2}) and (\ref{ham3}), and the
wave packets (\ref{pck3}), will then be the same if we can prove
that the difference between the integrals in (\ref{trm4}) and
(\ref{term5}) vanish for asymptotically large time. This amounts
to showing the following theorem whose proof is given in the
Appendix.
\newtheorem{main}{Theorem}
\begin{main}
\label{main} Let $f,g,h_\mu ,\psi,\psi '$ be defined as in
(\ref{trm4}) and (\ref{term5}). Then
\begin{equation}
\label{term6}
\lim_{t\rightarrow\infty} \int d^3 q \,d^3k [f(\qb+\kb)e^{-i\psi
t}- f(\qb ) e^{-i\psi 't}]g(\qb )q^\mu h_\mu(\kb )=0\,.
\end{equation}
\end{main}
This shows that the physics of scattering in scalar QED at large
times is fully described by the asymptotic Hamiltonian
(\ref{ham3}). The four point interaction vanishes and the three
point one can be described using the simple current (\ref{cur2}).
We have thus extended the results of KF to scalar QED and proved
the spin independence of the asymptotic dynamics in abelian gauge
theories.

\section[Conclusions]{Conclusions}
It is not necessary to \emph{assume} that the coupling constant
asymptotically switches off. As we have seen, one can, for
theories like massive $\phi^4$, prove that the asymptotic dynamics
is free or, for theories like QED, with rather more effort,
determine the form of this asymptotic interaction. The arguments
for determining the asymptotic properties of interactions in
quantum field theories, proposed by Kulish and
Fadeev~\cite{kulish:1970}, have been improved upon and made
applicable to a more general type of interaction, including four
point couplings. The principle refinement of our approach to the
asymptotic dynamics is that we  examine the asymptotic properties
of matrix elements corresponding to specific interactions rather
than considering operators. This has the advantage of requiring
only the machinery for the convergence of sequences of c-numbers
rather than the more elaborate needs of operator convergence.

In the case of $\phi^4$ theory it was found that, when the
incoming wave packet had test functions with non-overlapping
supports in velocity space, the asymptotic interaction Hamiltonian
is weakly vanishing. This condition on the test functions, of
having non-overlapping support in velocity space, is exactly that
which is required in the LSZ formalism and the construction of the
$S$-matrix. Our result is in complete agreement with perturbation
theory and shows why it works.

In the case of scalar QED we used our methods to show that the
matrix elements associated with the four point interaction term
are all asymptotically trivial which is again in line with the
results of perturbation theory. For the three point interaction
term of scalar QED, our methods show how the asymptotic dynamics
associated with the event of a charged particle emitting a photon
can be shown to be exactly the same as that of a charged particle
with known non-trivial asymptotic dynamics, and this conformed to
the approximation given by KF. The spin independence of this
result immediately translates to the fermionic theory.

What can we learn from this work about QED? Firstly that the
coupling does not \lq switch off\rq\ at large times. KF further showed that
this implies that the Lagrangian matter field does \emph{not}
asymptotically approach the free field of the plane wave expansion.
Rather there is a distortion factor which expresses itself in
perturbation theory in the branch cuts (instead of poles) in the matter
field two point function. They drew the conclusion from this that it is
not possible to describe charged particles in QED. Although this paper supports
the non-vanishing of the interaction, we feel that this last conclusion
is not justified. What one needs is to find the fields which do asymptotically
approach the plane wave expansion and can therefore be interpreted as
particles. (It is in fact clear from the start that the
Lagrangian matter cannot hope to do this since it is
not gauge invariant.) That such fields exist has been shown
elsewhere~\cite{Bagan:1999xx}
and that their Green's functions have a good pole structure has
been amply demonstrated, see,
e.g.,~\cite{Bagan:1997su,Bagan:1998kg,Bagan:1999yy}.
These are the physical fields which should be
identified with the charged particles seen in experiment.

We would like to suggest here three further areas for study:
massless QED is a theory with collinear divergences and as such a
playground for understanding QCD. The asymptotic dynamics of this
theory~\cite{Horan:1998im} requires further study, in particular
the physical asymptotic fields need to be constructed. Finite
temperature field theory is another area where infra-red
divergences are important, here, of course, the residual
asymptotic dynamics of zero temperature will be acerbated by
excitations from the heat bath. Finally in QCD confinement shows
that the interaction does not switch off and the strong
interaction between quarks and gluons is indeed supposed to grow
with the separation. The application of the methods of asymptotic
dynamics and the construction of physical fields at short
distances~\cite{Lavelle:1998dv,Lavelle:1997ty} could have
implications for jets production. It has though been demonstrated
that there is a topological obstruction to the construction of an
isolated quark or gluon~\cite{Lavelle:1997ty}, how this relates to
non-perturbative effects in the asymptotic dynamics of QCD is a
topic for future work.
\newpage

\no{\Large {\bf Acknowledgements}}
\newline
\newline
We would like to thank Emili Bagan and Izumi Tsutsui for
valuable discussions and suggestions during the course of this
work.
\appendix
\section[Appendix]{Appendix}

In this section we shall provide a proof of Theorem~\ref{main}.
This is split into two parts: First we shall show that the two
integrals
$$
 \int d^3 q \,d^3k  f(\qb )g(\qb )q^\mu h_\mu(\kb
) e^{-i\psi 't}\quad \mathrm{and}\quad\int d^3 q \,d^3k f(\qb+\kb
)g(\qb )q^\mu h_\mu(\kb ) e^{-i\psi 't} $$
are asymptotically equivalent. Then we shall show that the two
integrals
$$
 \int d^3 q \,d^3k  f(\qb+\kb )g(\qb )q^\mu h_\mu(\kb )
e^{-i\psi 't}\quad \mathrm{and}\quad\int d^3 q \,d^3k f(\qb+\kb
)g(\qb )q^\mu h_\mu(\kb ) e^{-i\psi t} $$
are asymptotically equivalent. Together, they give the proof of
the theorem. The first of these results is
\newtheorem{lem1}{Lemma}
\begin{lem1}
\label{lem1} Let $f,g,h_\mu ,\psi '$ be as defined in
(\ref{term5}). Then the two integrals
\begin{equation}
\label{lem1a} \int d^3 q \,d^3k  f(\qb )g(\qb )q^\mu h_\mu(\kb )
e^{-i\psi 't}\quad and\quad\int d^3 q \,d^3k f(\qb+\kb )g(\qb
)q^\mu h_\mu(\kb ) e^{-i\psi 't}
\end{equation}
have the same asymptotic limit, i.e., if ${\cal I}_t$ is defined
by
\begin{equation}
\label{ asm1}
{\cal I}_t= \int d^3 q\, d^3k \,[f(\qb +\kb )- f(\qb )]g(\qb
)q^\mu h_\mu(\kb ) e^{-i\psi 't}
\end{equation}
then ${\cal I}_t\rightarrow 0$ as $|t|\rightarrow \infty.$
\end{lem1}

\no {\bf Proof}\  We first take a fixed value of $\qb $ and
consider
\begin{equation}
\label{asm2}
{\cal I}_t(\qb )= \int  d^3k \, [f(\qb +\kb )- f(\qb )]q^\mu
h_\mu(\kb ) e^{-i\psi 't}\,.
\end{equation}
This integral is clearly well defined and there is a a positive
number $M$, say, such that $\int d^3k \,|q^\mu h_\mu(\kb )|<M\,.$

 Now given $\varepsilon >0$, choose a $\delta>0$ such that
$|f(\qb +\kb )- f(\qb )|<\frac{\varepsilon}{2M} $ if $|\kb |
<\delta\,.$ Let $U_1=\{ \kb : |\kb |<\delta \} $ and $U_2=\{ \kb :
|\kb |>\delta /2\}$  and let $\rho_1,\rho_2$ be a smooth partition
of unity subordinate to $U_1,U_2$ respectively. Write
\begin{eqnarray}
\label{part1}
{\cal I}_t^1(\qb )= \int  d^3k\, \rho_1(\kb )\left(f(\qb +\kb )-
f(\qb )\right)q^\mu h_\mu(\kb ) e^{-i\psi 't}\nonumber\\
&&\\
{\cal I}_t^2(\qb )= \int  d^3k \,\rho_2(\kb )\left(f(\qb +\kb )-
f(\qb )\right)q^\mu h_\mu(\kb ) e^{-i\psi ' t}\,.\nonumber
\end{eqnarray}
For the first of these we have $|{\cal I}_t^1(\qb )|<\varepsilon
/2$ by construction. For the second, the integrand  in ${\cal
I}_t^2(\qb )$ is defined on $U_2,$ which does not contain zero, so
that $\psi '$ is differentiable on the domain of integration
(i.e., $U_2$)\,. One can then apply the method of stationary phase
to this integral, using the partial derivatives of $k_i$, to show
that ${\cal I}_t^2(\qb )\rightarrow 0$ as $|t|\rightarrow
\infty\,.$

 We have shown that  ${\cal I}_t(\qb )\rightarrow 0$ as
$|t|\rightarrow \infty,$ for every $\qb$. Note that ${\cal
I}_t(\qb )$ satisfies the inequality $|{\cal I}_t(\qb )|\leq\int
d^3k\,|f(\qb+\kb)-f(\qb)|\,|q^\mu h_\mu(\kb )|$, and the latter is
in  ${\mathrm L}^1(\qb)$. The Lebesgue Dominated Convergence
Theorem \cite{Rudin:1970} may now be invoked to prove that ${\cal
I}_t \rightarrow 0$ as $|t|\rightarrow \infty,$ and this completes
the proof.  $\Box$

 Lemma~\ref{lem1} means that we have to show that the two
integrals
\begin{equation}
\label{lem1b} \int d^3 q \,d^3k f(\qb+\kb )g(\qb )q^\mu h_\mu(\kb
) e^{-i\psi 't}\quad\mathrm{and}\quad\int d^3 q\, d^3k
 f(\qb +\kb )g(\qb )q^\mu h_\mu(\kb ) e^{-i\psi t}
\end{equation}
have the same asymptotic limits, where $\psi,\psi '$ are defined
in (\ref{term5}) and (\ref{trm4}) respectively.

 Before proceeding with this, let us recall the form of
Taylor's theorem for a smooth function $u$, i.e.
\begin{eqnarray}
\label{taylor}
u(\qb +\kb )-u(\qb )-\partial _iu(\qb )k_i&=&R(\qb ,\kb )\nonumber
\\
\mathrm{and}\qquad \qquad\frac{|R(\qb ,\kb )|}{|\kb |}\rightarrow
0&\mathrm{as} &|\kb |\rightarrow 0\,.
\end{eqnarray}
Thus, for a fixed value of $\qb$ and  a given $\varepsilon >0,$
there is a $\delta
>0$ such that if $|\kb |<\delta$ then $|R(\qb ,\kb
)|/|\kb |<\varepsilon\,.$ If the choice of $\kb$ is restricted to
the set $\{\kb : |\kb |<1\},$ then $\exists\, t_0>0 $ such that
$\forall \,|t|>t_0,$ we have $|R(\qb ,\kb /t )|/|\kb /t
|<\varepsilon\,$ and from this we are able to conclude that $|t
R(\qb ,\kb /t )|<\varepsilon$ for all $|t|>t_0\,.$

 We shall now prove our final theorem.
\newtheorem{thm}[main]{Theorem}
\begin{thm}
\label{thm} Let $f,g,h_\mu ,\psi,\psi '$ be defined as in
(\ref{trm4}) and (\ref{term5}). If we define $I_t$ as
\begin{equation}
\label{asm3}
I_t= \int d^3 q \, d^3k \, f(\qb +\kb )g(\qb )q^\mu h_\mu(\kb )
(e^{-i\psi t}-e^{-i\psi 't})\,,
\end{equation}
then $I_t\rightarrow0$ as $|t|\rightarrow\infty.$
\end{thm}
 {\bf Proof}\  In the following discussions the value of the
functions $f,g,h_\mu$ in (\ref{asm2}) are not important and the
only property that is required of them is that they are test
functions. For convenience, therefore, we shall write the integral
as
\begin{equation}
\label{asm4}
I_t= \int d^3 q \,d^3k \, h(\qb ,\kb )(e^{-i\psi t}-e^{-i\psi
't})\,.
\end{equation}
with $h(\qb ,\kb )$ being a test function.

Take a fixed value of $\qb $ and write
\begin{eqnarray}
\label{asm5}
I_t(\qb )&=& \int d^3k \, h(\qb ,\kb )e^{-i\psi t}(1-e^{i(\psi -
\psi ')t}) \nonumber\\
&=&\int d\!\left(\frac{\kb }{t}\right) \, h\!\left( \qb ,\frac{\kb
}{t}\right) e^{-i\psi(\kb /t) t}(1-e^{i\Phi(\kb /t) t})\,.
\end{eqnarray}
As in Lemma~\ref{lem1}, we shall first show that $I_t(\qb )$ has a
zero asymptotic limit. In the latter integral we have changed the
variable from $\kb $ to $\kb /t$ so that  now we are writing $\psi
(\kb /t) =E_{\qb +\kb /t}-E_\qb -\omega_{(\kb /t)}$ and $\Phi(\kb
/t) =E_{\qb +\kb /t}-E_\qb-(\qb \cdot \kb) /(tE_\qb)\,.$ The
latter expression is in a form that will allow the use of Taylor's
theorem.

Let $U_1,U_2$ be the open cover of $ \R ^3$ given by $U_1=\{\kb
\,:\,|\kb |<1\},\,U_2=\{\kb \,:\,|\kb |>\frac{1}{2}\},$ and let
$\rho_1,\rho_2$ be a smooth partition of unity subordinate to this
cover. We write the second integral in (\ref{asm5}) as the sum of
two integrals by incorporating this partition, i.e.
\begin{eqnarray}
\label{asm6}
I^1_t(\qb )&=&\int d\!\left(\frac{\kb }{t}\right) \,\rho_1(\kb )\,
h\!\left( \qb ,\frac{\kb }{t}\right) e^{-i\psi(\kb /t)
t}(1-e^{i\Phi (\kb /t) t})\nonumber\\
&&\\
I^2_t(\qb )&=& \int d\!\left(\frac{\kb }{t}\right) \,\rho_2(\kb
)\, h\!\left( \qb ,\frac{\kb }{t}\right) e^{-i\psi(\kb/t)
t}(1-e^{i\Phi(\kb/t) t})\nonumber\\
&=&\int d\!\left(\frac{\kb }{t}\right) \,\rho_2(\kb ) h\!\left(
\qb ,\frac{\kb }{t}\right)( e^{-i\psi(\kb /t) t}-e^{-i\psi ' (\kb
/t)t}) \nonumber\,,
\end{eqnarray}
and we shall deal with $I_t^1(\qb )$ first.

 Due to the presence of $\rho_1,$ the integral  $I_t^1(\qb )$
is defined on $\{\kb :|\kb |<1\}$. Now given
$\varepsilon>0,\,\exists\, \delta'>0$ such that if
$|\theta|<\delta'$ then $|1-e^{i\theta}|<\varepsilon\,.$ From
Taylor's theorem, (see (\ref{taylor}) and the paragraph following
it) we can find a $t_0>0$ such that, for all $|\kb |<1,$ if
$|t|>t_0$ then $|\left(E_{\qb +\kb /t}-E_{\qb }-(\kb \cdot \qb)
/(tE_{\qb })\right)t|=|t R(\qb ,\kb /t)|<\delta'$, say. Then if
$|t|>t_0$, we have
\begin{equation}
\label{asm7}
|I_t^1(\qb )|\leq\int d\!\left(\frac{\kb
}{t}\right)\left|h\!\left( \qb ,\frac{\kb
}{t}\right)\right|\varepsilon=\varepsilon\int d^3k |h(\qb ,\kb
)|\,,
\end{equation}
where the latter integral is well defined, is independent of $t$
and has been obtained from the previous one by changing the
variables.

 The next step is to show that the asymptotic limit of $I_t^2(
\qb )$ is zero. For this we consider the last integral in
(\ref{asm6}) as the difference of the two obvious integrals given
by their exponential terms, i.e.
\begin{eqnarray}
\label{two}
I^2_1(t,\qb )&=&\int d\!\left( \frac{\kb }{t}\right) \,\rho_2( \kb
)\, h\!\left( \qb ,\frac{\kb }{t}\right)\, e^{-i\psi(\kb /t)t}
\nonumber
\\
&=&\int d^3 k \,\rho_2( \kb t)\, h( \qb ,\kb )\, e^{-i\psi( \kb)t
} \nonumber
\\
&& \\
I^2_2(t,\qb )&=&\int d\!\left(\frac{\kb }{t}\right) \,\rho_2(\kb )
\,h\!\left( \qb ,\frac{\kb }{t}\right)\,e^{-i\psi ' (\kb /t)t}
\nonumber
\\
&=&\int d^3 k \,\rho_2( \kb t)\, h( \qb ,\kb )\, e^{-i\psi ' ( \kb
)t } \nonumber \,,
\end{eqnarray}
with the final forms of $I^2_1,\,I^2_2$ being obtained by the
obvious change of variables.

 We shall first examine $I_1^2$ and we begin by changing the
variable for $\kb $ again, this time to $\kb = \omega \hat{k }$
with $|\hat{k }|=1$ and $\omega\geq0,$ i.e. polar coordinates. We
now have
\begin{equation}
\label{fst}
I^2_1(t,\qb )=\int d \hat{k }\,\int^{\infty}_0\omega^2 d\omega
\,\rho_2( \hat{k }\omega t)\, h( \qb ,\hat{k }\omega t )\,
e^{-i\psi( \hat{k }\omega )t
 }\,.
\end{equation}

Note that $\rho_2(\kb t )=\rho_2(\hat{k }\omega t )$ and this will
be zero for $0\leq\omega t\leq\frac{1}{2}\,.$ The exponent of the
integral in (\ref{fst}) is
$$
\psi(\hat{k }\omega )=E_{\qb +\hat{k }\omega }-E_{q}-\omega
\nonumber
$$
so that
\begin{equation}
\label{exp2}
\xi_1(\omega)\stackrel{\mathrm{def}}{=}\frac{\partial\psi}{\partial\omega}=\frac{\qb
\cdot\hat{k }+\omega}{E_{\qb +\hat{k }\omega }}-1\,.
\end{equation}
As $\omega\rightarrow 0$, $\xi_1(\omega)$ tends, uniformly in
$\hat{k },$ to $(\qb \cdot\hat{k }/E_{\qb })-1$, and since $|\qb
\cdot\hat{k }/E_{\qb }|\leq(|\qb |/E_\qb) <1,$ we have
$\xi_1(\omega)=\partial\psi/\partial\omega $ is bounded, uniformly
in $\hat{k }$, strictly away from $0$  in a neighbourhood of
$\omega=0\,.$ It is also easy to check that $\xi_1(\omega)$ is
non-zero for any finite value of $\omega\,.$

 We now have
\begin{eqnarray}
\label{stp1}
I^2_1(t,\qb )&=&\int d\hat{k }\,\int_0^\infty \omega^2
d\omega\,\rho_2(\hat{k }\omega t)\,h(\qb ,\hat{k }\omega
)\left(-\frac{1}{it\xi_1(\omega)}\right)\frac{\partial}{\partial
\omega}e^{-i\psi(\hat{k }\omega )t}\nonumber\\
&=&\frac{1}{it}\int d\hat{k }\,\int_0^\infty
d\omega\,\frac{\partial}{\partial\omega}\left(\rho_2(\hat{k
}\omega t)\,\omega^2h(\qb ,\hat{k }\omega
)\frac{1}{\xi_1(\omega)}\right)e^{-i\psi(\hat{k }\omega )t},
\end{eqnarray}
with the latter expression being obtained after integration by
parts in $\omega,$ and noting that the boundary terms vanish. This
can then be written as the sum of two integrals
\begin{eqnarray}
\label{stp2}
I^2_1(t,\qb )&=&\frac{1}{it}\int d\hat{k }\,\int_0^\infty
d\omega\,\rho_2(\hat{k }\omega t)\frac{\partial
}{\partial\omega}\left(\frac{\omega^2h(\qb ,\hat{k
}\omega)}{\xi_1(\omega)}\right)e^{-i\psi(\hat{k }\omega
)t}\nonumber\\
& &+\frac{1}{it}\int d\hat{k }\,\int_0^\infty
d\omega\,\left(\frac{\partial }{\partial\omega}\rho_2(\hat{k
}\omega t)\right)\left(\frac{\omega^2h(\qb ,\hat{k
}\omega)}{\xi_1(\omega)}\right)\, e^{-i\psi(\hat{k }\omega )t}\,.
\end{eqnarray}

If $v_1(\qb,\hat{k }\omega)\stackrel{\mathrm{def}}{=}\omega^2h(\qb
,\hat{k }\omega)/\xi_1(\omega)$ then $\partial v_1(\qb,\hat{k
}\omega)/\partial\omega$ is rapidly decreasing at infinity. The
first integral in (\ref{stp2}) can now be disposed of since it is
bounded by
 $(1/|t|)\int d\hat{k }\,\int_0^\infty d\omega\,\left|\partial v_1(\qb,\hat{k
}\omega)/\partial\omega\right|$ and since this is well defined,
vanishes  as $|t|\rightarrow\infty\,.$

Before dealing with the second integral in (\ref{stp2}) it is
worthwhile examining the properties of the derivatives of
$\rho_2.$ We can write this function as $\rho_2(\hat{k
}\omega)=\rho(\hat{k },\omega),$ emphasising the fact that
$\rho_2$ is a function of two variables. Now let us write
\begin{equation}
\label{rho} \varrho(\hat{k
}\omega)\stackrel{\mathrm{def}}{=}\frac{\partial}{\partial\omega}\rho_2(\hat{k
}\omega)=\partial_2\rho(\hat{k },\omega)
\end{equation}
 Since
$\rho_2$ is smooth and $\rho_2(\hat{k }\omega)=1$ for $\omega>1,$
the function $\varrho(\hat{k }\omega)$ is a smooth function which
has the important property that it is zero for $\omega>1,$ i.e.,
$\varrho$ has its support in $\{\omega:1/2\leq\omega\leq1\}.$

 The second integral in (\ref{stp2}), which we denote by ${\cal J}_1^t( \qb) ,$ is now
\begin{eqnarray}
\label{lim1}
{\cal J}_1^t( \qb) &=&\frac{1}{it}\int d\hat{k }\,\int_0^\infty
\omega^2d\omega\,\frac{\partial
}{\partial\omega}\left(\rho_2(\hat{k }\omega
t)\right)\,v_1(\qb,\hat{k }\omega) e^{-i\psi(\hat{k }\omega
)t}\nonumber\\
&=&\frac{1}{it}\int d\hat{k }\,\int_0^\infty\omega^2 d\omega
\,t\,\varrho(\hat{k }\omega t)\,v_1(\qb,\hat{k
}\omega)\,e^{-i\psi(\hat{k }\omega )t}\\
&=&\frac{1}{i}\int d\hat{k }\,\int_0^\infty \omega^2d\omega
\,\varrho(\hat{k }\omega t)\,v_1(\qb,\hat{k
}\omega)\,e^{-i\psi(\hat{k }\omega )t}\,.\nonumber
\end{eqnarray}
The last integral in (\ref{lim1}) is now written in a form which
is more convenient and which is obtained by the following changes
of variables. First replace $\hat{k }\omega$ by $\kb ,$ and then
replace $\kb $ by $\kb /t\,.$ The form of ${\cal J}_1^t( \qb)$
then changes to
\begin{equation}
\label{lim2}
{\cal J}_1^t( \qb)=\frac{1}{i}\int d\!\left(\frac{\kb
}{t}\right)\, \varrho (\kb )\, v_1\left(\qb ,\frac{\kb
}{t}\right)\,e^{-i\psi(\frac{\kb }{t} )t}
\end{equation}

 In order to deal with (\ref{lim2}) we shall first have to
examine the second integral in (\ref{two}). Applying the same
methods to $I^2_2$ as we have to $I^2_1$  one can easily show that
$I^2_2$ can also be written as the sum of two integrals, as in
(\ref{stp2}), but with the exponent $\psi$ replaced with $\psi '$
and $\xi_1$ replaced with $\xi_2,$ where
\begin{equation}
\label{exp3}
\xi_2(\omega)\stackrel{\mathrm{def}}{=}\frac{\partial \psi
'}{\partial\omega}(\hat{k }\omega)=\frac{\qb \cdot\hat{k
}}{E_q}-1\,.
\end{equation}
Thus
\begin{eqnarray}
\label{stp3}
I^2_2(t,\qb )&=&\frac{1}{it}\int d\hat{k }\,\int_0^\infty
d\omega\,\rho_2(\hat{k }\omega t)\frac{\partial
}{\partial\omega}\left(\frac{\omega^2 h(\qb ,\hat{k
}\omega)}{\xi_2(\omega)}\right)e^{-i\psi '(\hat{k }\omega
)t}\nonumber\\
& &+ \frac{1}{it}\int d\hat{k }\,\int_0^\infty\omega^2
d\omega\,\frac{\partial }{\partial\omega}(\rho_2(\hat{k }\omega
t))\left(\frac{h(\qb ,\hat{k }\omega)}{\xi_2(\omega)}\right)\,
e^{-i\psi '(\hat{k }\omega )t},
\end{eqnarray}
c.f. (\ref{stp2}). The function $\xi_2$ has similar properties to
$\xi_1$ and the first integral in (\ref{stp3}) vanishes
asymptotically in a similar fashion to the corresponding integral
in (\ref{stp2}). This leaves us with the latter integral in
(\ref{stp3}), which we shall write as ${\cal J}_2^t(\qb )\,.$
Using the same changes of variables, this can can be written as
\begin{equation}
\label{lim3}
{\cal J}_2^t( \qb)=\frac{1}{i}\int d^3\,\left(\frac{\kb
}{t}\right)\, \varrho (\kb )\, v_2\left(\qb ,\frac{\kb
}{t}\right)\,e^{-i\psi '(\frac{\kb }{t} )t}
\end{equation}

 The final objective is to show that ${\cal J}_1^t( \qb)-{\cal J}_2^t( \qb)$
 is asymptotically  vanishing and this can be achieved in
two steps.

 First define ${\cal J}_3^t( \qb)$ as
\begin{equation}
\label{lim4}
{\cal J}_3^t( \qb)=\frac{1}{i}\int d^3\,\left(\frac{\kb
}{t}\right)\, \varrho (\kb )\, v_1\left(\qb ,\frac{\kb
}{t}\right)\,e^{-i\psi '(\frac{\kb }{t} )t}\,.
\end{equation}
i.e., the $v_2$ in ${\cal J}_2^t( \qb)$ is replaced by $v_1.$ Then
it is straightforward to prove, along the lines of
Lemma~\ref{lem1}, that, for all $\qb, $ we have ${\cal J}_2^t(
\qb)-{\cal J}_3^t( \qb)\rightarrow0$ as $|t|\rightarrow\infty.$

 Secondly, we must show that ${\cal J}_1^t( \qb)-{\cal J}_3^t( \qb)\rightarrow0$
 as $|t|\rightarrow\infty.$ This is also
straightforward and can be fashioned along the lines of the proof
that $I_t^1(\qb)\rightarrow0$ as $|t|\rightarrow\infty$ (see
(\ref{asm6}) and the paragraph that follows it.) We omit the
details.

 This proves that, for every $\qb,$  $I_t(\qb)\rightarrow0$ as
$|t|\rightarrow \infty$ (see(\ref{asm5})). The Lebesgue Dominated
Convergence Theorem now implies that $I_t\rightarrow0$ as
$|t|\rightarrow\infty,$ as required. $\Box$

\end{document}